
\magnification=1200
\hsize 15true cm \hoffset=0.5true cm
\vsize 23true cm
\baselineskip=15pt

\font\grande=cmr10 scaled \magstep4
\font\medio=cmr10 scaled \magstep2
\outer\def\beginsection#1\par{\medbreak\bigskip
      \message{#1}\leftline{\bf#1}\nobreak\medskip\vskip-\parskip
      \noindent}

\def \me {\buildrel <\over \sim}

\def \pa {\partial}
\def \ra {\rightarrow}
\def \pr {\prime}

\def \ti {\tilde}

\def \b {\beta}
\def \a {\alpha}

\def \Ga {\Gamma}
\def \ga {\gamma}

\def \da {\delta}

\def \om {\omega}
\def \Om {\Omega}
\def \noi {\noindent}

\def\sqr#1#2{{\vcenter{\hrule height.#2pt\hbox{\vrule width.#2pt
height#1pt \kern#1pt\vrule width.#2pt}\hrule height.#2pt}}}
\def\square{\mathchoice\sqr34\sqr34\sqr{2.1}3\sqr{1.5}3}
\def\lsim{\mathrel{\rlap{\lower4pt\hbox{\hskip1pt$\sim$}}
    \raise1pt\hbox{$<$}}}         
\def\gsim{\mathrel{\rlap{\lower4pt\hbox{\hskip1pt$\sim$}}
    \raise1pt\hbox{$>$}}}         

\line{\hfil DFTT-58/92}
\vskip 3truecm
\centerline {\grande Dilaton Contributions To The}
\vskip 1 true cm
\centerline{\grande Cosmic Gravitational Wave Background}
\vskip 1 cm
\centerline{M.Gasperini and M.Giovannini}
\centerline{\it Dipartimento di Fisica Teorica dell'Universit\`a,}
\centerline{\it Via P.Giuria 1, 10125 Torino, Italy,}
\centerline{\it and}
\centerline{\it Istituto Nazionale di Fisica Nucleare, Sezione di Torino}
\vskip 2 cm
\centerline{\medio Abstract}
\noindent
We consider the cosmological amplification of a metric perturbation
propagating in a higher-dimensional Brans-Dicke background, including a
non trivial dilaton evolution. We discuss the properties of the spectral
energy density of the produced gravitons (as well as of the associated
squeezing parameter), and we show that the present observational bounds
on the graviton spectrum provide significant information on the
dynamical evolution of the early Universe.
\vskip 2 truecm
\noi
-------------------------------------
\vskip 1 truecm
To appear in {\bf Phys.Rev.D}
\vfill\eject

{\bf 1. Introduction}

It is well known that the transition from a primordial inflationary phase
to a decelerated one, typical of our present cosmological evolution, is
associated with the production of a cosmic background of relic gravity
waves$^{1-6}$.
The spectral distribution of their energy density may provide direct
information on the very early history of our Universe, and can be used,
in particular, to reconstruct the time dependence of the Hubble
parameter$^{7}$.

Deflation, however, is not the only violent process typical of primordial
evolution
able to amplify a metric fluctuation. Although less known (or
less studied, up to now at least), it is a fact that
gravitons can be produced from
the vacuum also as a consequence of a phase of dynamical dimensional
reduction$^{8,9}$, in which a given number of "internal" dimensions
shrink down to a final compactification scale.
Another possible process which may lead to a cosmological graviton
production, and which (to our knowledge) has not yet been discussed
in the literature, is the time variation of the effective
 gravitational coupling constant $G$.

The main purpose of this paper is to compute the expected spectrum of
 the cosmic gravitons background, by including both the contributions of
dimensional reduction and of $\dot G$ among the possible sources
 (besides inflation), and by using a Brans-Dicke-like graviton-dilaton
coupling as a dynamical model of variable $G$. We are led to
this choice, in particular, by the models of early
universe evolution based on the low energy string effective action $^{10
-12}$, which suggest that the
standard radiation-dominated cosmology is preceeded
by a dual, "string-driven" phase, in which the effective gravitational
 coupling changes just because of the time-dependence of the
 dilaton background. The possibility of
looking for tracks of such a string phase in the
properties of the cosmic graviton spectrum provides indeed one
of the main motivations of the present work.

The paper is organized as follows.
In Section II we deduce the linearized equation for a gravitational wave
perturbation in a Brans-Dicke background, and, in Section III, this
equation
is used to compute the spectral distribution of the gravitons, produced
by the cosmological background transitions.
We shall take into account the dilaton-driven variation of $G$ in a
higher dimensional framework in which also the scale of the internal
spatial
dimensions is allowed to vary, and in which the matter-dominated and
radiation-dominated evolution of the external space follows a phase of
accelerated (i.e. inflationary) expansion.
The squeezing parameter$^{13}$ corresponding to this scenario will be
given in Section IV.

In Section V the present bounds on the energy density distribution
of the relic gravitons are used to obtain information, and constraints,
on the value of the curvature scale at the transition between
the inflationary and the radiation-dominated era, versus the parameters
characterizing the background kinematics.
The predictions of some string-inspired cosmological models (and of
related Kaluza-Klein scenarios) will be compared with these bounds in
Section VI.
The main conclusions of this paper will be finally summarized in
Section VII.

\vskip 2cm

{\bf 2. Gravitational perturbations of a Brans-Dicke background.}

The starting point to discuss the production of
gravitons, induced by a
cosmological background transition, is the linearized wave equation for
a gravitational perturbation propagating freely in the given background.
In order to include the effect of a changing gravitational coupling,
such an equation will be obtained by perturbing (at fixed sources)
the Brans-Dicke field equations, around a background configuration
which includes a
time-dependent dilaton field.

It should be perhaps recalled that, in a general relativity context
and in a spatially flat
Friedmann-Robertson-Walker manifold, a
gravity-wave perturbation  obeys the same equation  as a
minimally coupled massless scalar field$^{1,14,15}$.
In a Brans-Dicke context, however, the graviton wave equation is
different  from the covariant Klein-Gordon equation,
because the gravitational perturbations are coupled not
only to the background metric tensor, but also to the scalar dilaton
background $\phi (t)$ representing the $G$ variation.

Our background field dynamics is assumed to be described, in $D$
dimensions, by the following
scalar-tensor action
$$
S=-{1\over {16\pi G}}\int d^{D}x~e^{-\phi}~\sqrt{|g|}(R- \om g^{\mu \nu}
\pa_{\mu} \pa_{\nu}\phi ) + S_{m} \eqno(2.1)
$$
where $\om$ is the usual Brans-Dicke parameter, and $S_{m}$ represent the
possible contribution of matter sources, with $\sqrt{|g|}T_{\mu \nu}=2
\delta S_{m}/\delta g^{\mu \nu}$.
The variation of this action with respect to $\phi$ provide the dilaton
equation
$$
R+ \om (\nabla \phi)^{2}- 2\om\square\phi=0 \eqno(2.2)
$$
where $\nabla$ denotes the Riemann covariant derivative, and $\square=g^{
\mu \nu}\nabla_{\mu} \nabla_{\nu}$.
The variation with respect to $g_{\mu \nu}$, combined with (2.2),
provides the equation
$$
R_{\mu}\,^{\nu}+ {\nabla_{\mu}}{\nabla^{\nu}}\phi +(\om +1)[\delta_{\mu}^
{\nu}((\nabla\phi)^{2}-\square\phi) -\nabla_{\mu}\phi\nabla^{\nu}\phi]=
8\pi G_{D}e^{\phi}T_{\mu}\,^{\nu} \eqno(2.3)
$$
Note that we are using an exponential parametrization for the dilaton
field, to make contact with the string cosmology models.
For $\om=\infty$, $\phi=const$ we recover general relativity, while
for $\om=-1$ eq (2.1) reduces indeed to the truncated low energy
string effective action with phenomenological matter sources$^{10-12}$.

The free linearized wave equation for a metric fluctuation
$\delta g_{\mu \nu}=h_{\mu \nu}$ is now obtained by perturbing
eqs.(2.2) and (2.3), keeping all sources (dilaton included) fixed,
$\delta T_{\mu}^{\nu}=0=\delta\phi$.
It should be stressed that we have not explicitly included in
the action a possible dilaton potential term, $V(\phi)$, as
its contribution to the perturbation is vanishing for $\delta\phi=0$.
It is true that, in a class of duality-symmetric string cosmological
models$^{10-12,16}$, the dilaton self interactions may also occur trough
a coupling to the metric, and lead to
a two loop potential of the form $V=V_0
[\exp{(2\phi-2\ln{\sqrt{|g|})}}]$,
for which $\delta V\propto V\delta g$.
This potential, however, is expected to affect in a significant way only
the transition region between the inflationary and the radiation
dominated regime$^{12}$.
Its contribution to the perturbation equations may then be neglected
for the purpose of this paper where, as discussed in the following
section, we shall evaluate the graviton spectrum in the  "sudden"
approximation, namely in the approximation
in which the kinematic details of the transition regime
are ignored, and the rapid exponential decay of the high frequency tail
of the spectrum is replaced by a suitable high frequency cutoff.

We perform then the transformation $g_{\mu\nu} \ra g_{\mu\nu}+\da
g_{\mu\nu}$, with
$$
\delta g_{\mu \nu}= h_{\mu \nu}~~~~~~~~,~~~~~~~~
\delta \phi=0=\delta T_{\mu}\,^{\nu} \eqno(2.4)
$$
By neglecting corrections of order higher than first in $h_{\mu\nu}$
(so that, for instance, $\delta g^{\mu \nu}=-h^{\mu \nu}$), we are led to
the following variational expressions
$$
\delta(\nabla \phi)^{2}= -h^{\a \b}\pa_{\a}\phi \pa_{\b}\phi
$$
$$
\delta(\square \phi)=-h^{\mu \nu} \nabla_{\mu} \nabla_{\nu}\phi -
g^{\mu \nu} \pa_{\a} \phi \delta \Ga_{\mu \nu}^{\a}
$$
$$
\delta R= - h^{\mu \nu}R_{\mu \nu}+ g^{\mu \nu}\delta R_{\mu \nu}
$$
$$
\delta(\nabla_{\mu} \nabla^{\nu}\phi)=-h^{\nu
\a}\nabla_{\mu}\nabla_{\a}\phi -g^{\nu \a} \pa_{\b}\phi \delta \Ga_{\mu
\a}^{\b}
$$
$$
\delta(\nabla_{\mu}\phi\nabla^{\nu}\phi)=-h^{\nu\a}\pa_{\mu}\phi \pa_{\a}
 \phi \eqno(2.5)
$$
Here
$$
\delta \Ga_{\mu \nu}^{\a}={1\over 2}g^{\a \b}(\nabla_{\mu}h_{\nu \b}+
\nabla_{\nu}h_{\mu \b}- \nabla_{\b} h_{\mu \nu}) \eqno(2.6)
$$
and $\delta R_{\mu \nu}$ is the linearized expression for$R_{\mu \nu}(
\delta g)$ (note that all covariant derivatives, as well as all
operations of raising index on $h_{\mu\nu}$, are now to be understood as
performed with the help of
the background metric $g_{\mu \nu}$).

We choose, in particular, a time dependent background with $\phi= \phi(t
)$, and a homogeneous diagonal metric describing a general situation of
dimensional decoupling , in which $d$
dimensions expand with scale factor
$a(t)$, and $n$ dimensions contract with scale factor $b(t)$.
In a synchronous frame
$$
g_{00}=1~~,~~g_{ij}=-a^{2}(t)\ga_{ij}(x)~~,~~g_{ab}=-b^{2}(t)\ga_{ab}(y)
$$
$$
g_{0 \mu}=0= g_{ia}~~~~ ,~~~~ \phi=\phi(t) \eqno(2.7)
$$
(conventions: $\mu, \nu=1,...,D=d+n+1$; $ i,j=1,...,d$;
$ a,b=1,...,n$; $t$ is
the cosmic time coordinate, and $\ga_{ij}$, $\ga_{ab}$ are the metric
tensors of two maximally symmetric euclidean manifolds, parametrized
respectively by "internal" and "external" coordinates $\{x^{i}\}$ and
$\{y^a\}$).
We are interested moreover, in a pure tensor gravitational perturbation,
decoupled from sources, representing a gravitational wave propagating in
the $d$-dimensional
external space, such that $h_{\mu \nu}=h_{\mu \nu}(x,t
) $, $h_{0 \mu}=0=h_{a\mu}$, and which satisfies the transverse,
traceless gauge condition
$$
g^{\mu \nu}h_{\mu \nu}=0=\nabla_{\nu}h_{\mu}\,^{\nu} \eqno(2.8)
$$
In this case we have, for the background (2.7),
$$
\delta R=0~~,~~\delta \Ga_{\mu \nu}^{0}= -{1\over 2}\dot h_{\mu \nu}
\eqno(2.9)
$$
(a dot denotes derivative with respect to t).
The perturbation of eq.(2.2) is thus trivially satisfied, while
the perturbation of eq.(2.3) provides for $h_{\mu \nu}$ the
linearized wave equation
$$
\delta R_{\mu}\,^{\nu}+{1\over 2}{\dot\phi}\dot h_{\mu \a}g^{\nu \a}-
h^{\nu\a}\nabla_{\mu}\nabla_{\a} \phi = 0   \eqno(2.10)
$$
which, being $\om$ independent, is remarkably the same for all Brans-
Dicke  models.

The non-vanishing components of the background Ricci tensor, for the
metric (2.7), are given by
$$
R_{0}\,^{0}=-d(\dot H+H^{2})-n(\dot F + F^{2})
$$
$$
R_{i}\,^{j}=-{1\over a^{2}} \ti R_{i}\,^{j}(\ga(x))-
\delta_{i}^{j}(dH^{2}+
\dot H + nHF)
$$
$$
R_{a}\,^{b}=-{1\over b^{2}}\ti R_{a}\,^{b}(\ga(y))-
\delta_{a}^{b}(nF^{2} +
\dot F +dHF) \eqno(2.11)
$$
where $H={\dot a}/ a$, $ F={{\dot b}/ b}$ and $\ti R(\ga)$
denotes the Ricci tensor for the n dimensional euclidean spaces
computed from the metrics $\ga_{ij}(x)$ and $\ga_{a b}(y)$.
By using the relations
$$
\dot g_{ij}=2Hg_{ij}~~~~,~~~\dot g^{ij}=-2Hg^{ij} \eqno(2.12)
$$
one obtains$^{14,17}$, to the first order in $\delta g_{ij}=h_{ij}$,
$$
\delta (\dot g_{i}\,^{j})\equiv \delta(g^{jk}\dot g_{ik})=\dot h_{i}\,
^{j}
\equiv (g^{jk} h_{ik})^{.} \eqno(2.13)
$$
It is thus simple to show (in the gauge $g^{ij}h_{ij}=0$ ) that
$$
\delta H={1\over 2d}~\delta(g^{ik}\dot g_{ik})=0=\delta \dot H
$$
$$
\delta H^{2}={1\over 4d^{2}}~\delta(g^{ik}\dot g_{ik})^{2}=0
$$
$$
\delta(H \delta_{i}\,^{j})={1\over 2}\dot h_{i}\,^{j}
$$
$$
\delta(\dot H \delta_{j}\,^{i})={1\over 2}\ddot h_{i}\,^{j}
$$
$$
\delta(H^{2}\delta_{i}\,^{j})={1\over 2}H\dot h_{i}\,^{j} \eqno(2.14)
$$
(the corresponding perturbations of the F terms are all vanishing, since
$\delta g_{a b}=0$).
Therefore
$$
\delta R_{0}\,^{0}=0=\delta R_{a}\,^{b}
$$
$$
\delta R_{i}\,^{j}= -\delta({\ti R_{i}\,^{j}\over a^{2}})-
 {d\over 2}H\dot h_{i}^{j}
 - {1\over 2}\ddot h_{i}\,^{j} -{n\over 2}F\dot h_{i}\,^{j}
\eqno(2.15)
$$

We shall consider, in particular, a flat euclidean metric $\ga_{ik}=
\delta_{ik}$, so that $\Ga_{ij}^{k}(x)=0=\ti R_{i}\,^{j}(\ga)$.
The gauge condition ${\ti\nabla}(\ga)h_{i}\,^{j}=0$ reduces to
$\pa_{j}h_{i}\,^{j}=0$, and implies$^{14,17}$:
$$
\delta {\ti R}_{i}\,^{j}=-{1\over 2}\nabla^{2}h_{i}\,^{j} \eqno(2.16)
$$
with $\nabla^{2}=\delta^{ij}\pa_{i}\pa_{j}$.
We thus recover the usual result
$$
\delta R_{i}\,^{j}=-{1\over 2}[\ddot h_{i}\,^{j}+(dH+nF)\dot h_{i}\,^{j}-
{1\over a^2}\nabla^2 h_{i}\,^{j}]\equiv -{1\over 2}\square h_{i}\,^{j}
\eqno(2.17)
$$
valid whenever the background is isotropic in the polarization plane,
orthogonal to the direction of propagation of the wave$^{18}$.

On the other hand we have, for the background (2.7),
$$
\nabla_{i}\nabla_{j}\phi={1\over 2}\dot\phi\dot g_{ij} \eqno(2.18)
$$
Moreover, by using eq.(2.12),
$$
g^{jk}\dot h_{ik}-h^{jk}\dot g_{ik}=\dot h_{i}\,^{j}   \eqno(2.19)
$$
The linearised wave equation (2.10) thus reduces to
$$
\square h_{i}\,^{j}-\dot\phi\dot h_{i}\,^{j}=0   \eqno(2.20)
$$
and, in terms of the eigenstates of the Laplace operator,
$$
\nabla^{2}h_{i}\,^{j}(k)=-k^{2}h_{i}\,^{j}(k) \eqno(2.21)
$$
it takes the form
$$
\ddot h_{i}\,^{j}+(dH+nF-\dot\phi)\dot h_{i}\,^{j} +
 ({k\over a})^{2}h_{i}\,^{j}=0  \eqno(2.22)
$$

For later applications, it is convenient to rewrite this equation in
term of the conformal time coordinate $\eta$, defined by $dt/d\eta=a$.
Denoting with a prime the differentiation with respect to $\eta$, and
defining
$$
\psi_{i}\,^{j}=h_{i}\,^{j}a^{(d-1)/2}b^{n/2}e^{-\phi/2} \eqno
(2.23)
$$
we get finally, from eq.(2.22), that each polarization mode $\psi_{i}^{j}
(k)$ must satisfy the equation:
$$
\psi^{\pr\pr} + (k^{2}-V)\psi=0 \eqno(2.24)
$$
where

$$
V(\eta)={(d-1)\over 2}{a^{\pr\pr}\over a} + {n\over 2}{b^{\pr\pr}\over
b}
-{\phi^{\pr\pr}\over 2}+ {1\over 4}(d-1)(d-3)({a^{\pr}\over a})^{2}+
$$
$$
{1\over 4}n(n-2)({b^{\pr}\over b})^{2}+ {1\over 4}{\phi^{\pr}}^{2}
+{1\over 2}n(d-1){a^{\pr}b^{\pr}\over ab}-{1\over 2}(d-1){a^{\pr}\over
a}\phi^{\pr}-{n\over 2}{b^{\pr}\over b}{\phi^{\pr}} \eqno(2.25)
$$

This effective potential generalizes to a  higher  number of
dimensions the four-dimensional equation, used by Grishchuk
and collaborators$^{1,3,7}$, to study the cosmological
amplification of the quantum fluctuations of the metric tensor.
In addition, it takes into account the coupling of the metric
perturbations to a possible time variation of the gravitational coupling
constant ($\phi^{\pr}\not=0$ ), and to a possible variation of the scale
of n "internal" compactified dimensions ($b^{\pr}\not=0$). It may be
interesting to note that this potential can also be expressed in terms
of the scale factors only, by eliminating the explicit dilaton
dependence through the background equation (2.2), which implies
$$
-{\phi^{\pr\pr}\over 2}+
 {1\over 4}{\phi^{\pr}}^{2}
-{1\over 2}(d-1){a^{\pr}\over
a}\phi^{\pr}-{n\over 2}{b^{\pr}\over b}{\phi^{\pr}}
=
$$
$$
{1\over 4 \om}[2d{a^{\pr\pr}\over a}+2n{b^{\pr\pr}\over b}
+d(d-3)({a^\pr \over a})^2 +n(n-1)({b^\pr \over b})^2
+2n(d-1){a^\pr b^\pr \over ab}] \eqno(2.26)
$$
In this way one can re-introduce the $\om$-dependence which is otherwise
hidden in the particular choice of the dilaton background. For the
purpose of this paper, however, it will be more convenient to work
directly with the form (2.25) of the potential, in which $\phi$ appears
explicitly.
\vskip 2true cm
\centerline{\bf 3. Parametrization of the graviton spectrum}
\centerline{\bf for a general model of background evolution}
\vskip 0.5 true cm
As discussed in the previous section, the present day background of
cosmic gravitational waves may include, among its sources, not only a
metric transition (deflation, dynamical dimensional reduction), but also
a dilaton transition between two (or more) regimes with different
gravitational coupling.

In order to take all these contribution into account, we shall consider
the background metric of eq.(2.7) (with flat maximally symmetric
subspaces $\ga_{ij}=\delta_{ij}$, $\ga_{ab}=\delta_{ab}$), starting with
an initial configuration in which, for $\eta< -\eta_{1}$, $d$ dimensions
inflate with scale factor $a(\eta)$, $n$ dimensions shrink with scale
factor $b(\eta)$, and the dilaton coupling is growing according to
$$
a\sim\eta^{-\a}~~~,~~~b\sim\eta^{\b}~~~,~~~\phi\sim\ga\ln a~~~,~~~\eta<-
\eta_{1} \eqno(3.1)
$$
(note that in this equation $\eta$ ranges over negative values,
so that $\a$, $\b$ and $\ga$ are
all positive).
We shall assume that this phase is followed, at $\eta=-\eta_{1}$ and
$\eta=\eta_{2}$ respectively, by the standard radiation-dominated and
matter-dominated expansion of three spatial dimensions.
During these two last epochs,
however, the gravitational coupling and the
compactification scale of the possible additional $n_{1}$ internal
dimensions are not assumed to be frozen, but they are allowed to vary as
$$
a\sim\eta~~~,~~b\sim\eta^{-\b_{1}}~~~,~~~\phi\sim\ga_{1}\ln a~~~,
{}~~~-\eta_{1}<\eta<\eta_{2}
$$
$$
a\sim\eta^{2}~~~,~~~b\sim\eta^{-\b_{2}}~~~,~~~\phi\sim\ga_{2}\ln a~~~,
{}~~~0<\eta_{2}<\eta \eqno(3.2)
$$

According to this model of background evolution, the effective potential
(2.25) becomes
$$
V(\eta)={1\over {4\eta^{2}}}\left[ [\a(d-1-\ga)-n\b+1]^{2}-1
\right]~~~,~~~\eta<-\eta_{1}
$$
$$
V(\eta)={1\over {4\eta^{2}}}[(n_{1}\b_{1}+\ga_{1}-1)^{2}-1]]~~~,
{}~~~-\eta_{1
}<\eta<\eta_{2}
$$
$$
V(\eta)={1\over {4\eta^{2}}}[(n_{1}\b_{1}+2\ga_{2}-3)^{2}-1]~~~,
{}~~~\eta_{2}<\eta \eqno(3.3)
$$
(note that it goes to zero as $\eta\to\pm\infty$).
A particular solution of eq.(2.24) for $\psi(k)$ can thus be written in
terms of the first and the second kind Hankel functions $H^{(1)}$ and $H^{
(2)}$ (we follow the notation of Ref.(19)),
$\psi(k,\eta)\sim\eta^{1\over 2
}H_{\nu}^{(2,1)}(k\eta)$, which correspond to free
oscillating modes  in the
$|\eta|\to\infty$ limit, as
$\eta^{1\over 2}H^{(2,1)}(k\eta)\to {e^{\mp ik\eta}/ \sqrt{k}}$
(the minus and plus sign corresponds, respectively, to $H^{(2)}$ and
$H^{(1)}$).

The effective potential barrier (3.3) leads to an amplification of the
gravitational perturbations or, equivalently, to a graviton
production from the

\noi
vacuum$^{2,3,5-8}$.
Indeed, starting with incoming modes  which are of positive frequency
with respect to the vacuum at the left of the barrier
($\eta\to -\infty$),
one has in general, for $\eta\to +\infty$, a linear combination of
modes which are of positive and negative frequency, with respect to the
vacuum at the right of the barrier. The superposition coefficients
$c_{\pm}(k)$ define the Bogoliubov transformation$^{20}$ connecting the
"left" and "right" vacuum, and determine the spectral distribution of
the produced gravitons.

By assuming, in our case, the "in" states of the
gravitational field correspond to the Bunch-Davies "conformal" vacuum$^
{5,6,20}$, we can write the general solution of eq.(2.24), for each mode
$\psi(k)$, in the three temporal regions as follows:
$$
\psi_{I}(k)=C\eta^{1\over 2}H_{\nu}^{(2)}~~~~~~~~,~~~~~
{}~~~~~~~\eta<-\eta_{1}
$$
$$
\psi_{II}(k)=\eta^{1\over 2}[A_{+}H_{\mu}^{(2)}(k\eta)+ A_{-}H_{\mu}^{(1
)}(k\eta)]~~~~,~~~~-\eta_{1}<\eta<\eta_{2}
$$
$$
\psi_{III}(k)=\eta^{1\over 2}[B_{+}H_{\sigma}^{(2)}(k\eta)+B_{-
}H_{\sigma}^{(1)}(k\eta)]~~~,~~~\eta>\eta_{2} \eqno(3.4)
$$
where
$$
\nu={1\over 2}[\a(d-1-\ga)-n\b+1]
$$
$$
\mu={1\over 2}(n_{1}\b_{1}+\ga_{1}-1)
$$
$$
\sigma={1\over 2}(n_{1}\b_{2}+2\ga_{2}-3) \eqno(3.5)
$$
and $C$ is a normalization constant.
The Bogoliubov coefficients are given
by $c_{\pm}(k)={B_{\pm}/ C}$, and can be fixed by the four conditions
obtained matching $\psi$ and $\psi^{\pr}$ at $\eta=-\eta_{1}$ and $\eta=
\eta_{2}$.

The coefficients determined in this "sudden"
approximation lead,
however, to an ultraviolet divergence of the energy density of the
produced particles. The reason is that, for modes of comoving frequency
$k^{2}$ higher than the height of the potential barrier, the sudden
approximation is no longer adequate, and the mixing coefficients should
be computed by replacing the potential step with a smooth transition
of $V(\eta)$. In this way one finds, indeed, that the mixing of the
modes with $k>|V|^{1\over 2}$ is exponentially suppressed with respect
to the other modes$^{8,20,21}$, and the ultraviolet divergence is
avoided. In this paper, however, we are mainly interested in the general
behaviour of the spectral distribution, and not in the details of the
transition regime. We shall completely neglect, therefore, the frequency
mixing of modes which never "hit" the potential barrier, by putting, for
such modes, $c_{+}(k)\simeq 1$, $c_{-}(k)\simeq 0$.
This replaces the exponential decay of the high frequency side of the
spectrum with a cutoff, at an appropriate frequency $k\simeq|V|^{1\over
2}$.

Our potential barrier (3.3) has two steps, which satisfy $V(\eta_{1})
\simeq\eta_{1}^{-2}\gg\eta_{2}^{-2}\simeq V(\eta_{2})$ (for realistic
values of the parameters). The propagation of modes with $\eta_{2}^{-1}<
k<\eta_{1}^{-1}$ will thus be affected, in our approximation, only by the
first background transition at $\eta=\eta_{1}$. In this frequency band,
the Bogoliubov coefficients are then defined by $c_{\pm}={A_{\pm}/ C
}$; by matching $\psi_{I}$, $\psi_{II}$ and their first derivatives  at
$\eta=\eta_1$, and by using the small argument limit of the Hankel
functions, we obtain (for $k\eta_{1}<1$)
$$
c_{\pm}={1\over 2}\left[\ga({k\eta_1\over 2})^{\nu - \mu} \pm \ga^{-1}(
{k\eta_1\over 2})^{\mu - \nu}\right] \eqno(3.6)
$$
(here $\ga={\Ga(\mu)/ \Ga(\nu)}$, where $\Ga$ is the Euler function,
and we have supposed $\mu>0$, $\nu>0$ when performing the
$k\to 0$ limit).

These coefficients
satisfy correctly the Bogoliubov normalization condition, $|c_{+}|^{2}-
|c_{-}|^{2}=1$, and have been obtained in a more particular case$^{15}$,
and also with a different procedure$^{6,22}$, in previous papers.
For $k\eta_{1} <1$, we shall keep the dominant term only, ignoring
corrections to the sudden approximation near th maximum frequency $k_{1}
=\eta_{1}^{-1}$, and neglecting also numerical factors of order unity,
which depend on the model of background evolution (continuity of the
scale factors and of the dilaton at the transition time), and
which do not
affect the qualitative behaviour of the spectrum. In the rest of the
paper, therefore, we shall use the expression
$$
|c_{-}(k)|=(k\eta_{1})^{-|\mu - \nu|}~~~~~,~~~~~k_{2}<k<k_{1} \eqno(3.7)
$$
where $k_{2}={1/ \eta_{2}}$ is the frequency corresponding to the
height of the barrier $V(\eta_{2})$.

Lower frequency modes, $k<k_{2}$, are affected also by the second
background transition, at $\eta=\eta_{2}$, from the radiation to the
matter-dominated regime$^{3,5,6}$. In this frequency sector the
Bogoliubov coefficients are given by $c_{\pm}={B_{\pm}/ C}$, and the
matching condition provide, for $k\eta_{2}<1$,
$$
c_{\pm}(k)={1\over 2}\left[\ga_{1}({k\eta_{1}\over 2})^{\nu- \mu}\ga_{2
}({k\eta_{2}\over 2})^{\mu- \sigma} \pm \ga_{1}^{-1}({k\eta_{1}\over 2})
^{\mu- \nu}\ga_{2}^{-1}({k\eta_{2}\over 2})^{\sigma- \mu}\right]
\eqno(3.8)
$$
where $\ga_{2}={\Ga(\sigma)/ \Ga(\mu)}$ for $\mu >0$ and $\sigma >0$.
It may be useful to note that the expression can be easily generalized ,
by performing the product of $n$ Bogoliubov transformations, to the case
of n background transitions, at $\eta=\eta_{i}$, between the mode
solutions $H_{\nu_{i}}$ and $H_{\nu_{i+1}}$, with $i=1,2,...,n$.
One finds, in general$^{22}$,
$$
c_{\pm}^{(n)}={N\over 2}\prod_{i=1}^{n}\left[\ga_{i}({k\eta_{i}\over 2})
^{\nu_{i} - \nu_{i+1}} \pm \ga_{i}^{-1}({k\eta_{i}\over 2})^{\nu_{i+1}-
\nu_{i}}\right] \eqno(3.9)
$$
where $\ga_{i}$ are numerical factors of order unity, and $N^{\ast}=N^{-1
}$ is an overall constant phase factor.

In  order to keep only the dominant term of eq.(3.8), for $k< k_{2
}\ll k_{1}$, we have  to note first of all that the phenomenological
constraints on the time variation of the fundamental constant (including
G), during the matter and radiation-dominated era, imply $\sigma- \mu<0$
(see Sec.5). If $\mu- \nu<0$ (as
seems to be indeed the case for all the appropriate
models of background evolution, see Sec.6), the second term on the r.h.s.
of eq.(3.8) is the dominant one. If, on the contrary, $\mu- \nu>0$,
then the first term is dominant (for realistic values of $\eta_{1}$ and
$\eta_{2}$). We shall thus use, for the graviton production at low
frequencies,
$$
|c_{-}(k)|\simeq (k\eta_{1})^{-|\mu- \nu|}(k\eta_{2})^{\mp|\sigma- \mu|}
{}~~~~,~~~~k_{0}<k<k_{2}  \eqno(3.10)
$$
where the $-(+)$ sign refers to $\mu - \nu<0$ ($>0$), and $k_{0}$ is the
minimal amplified frequency$^{3,5}$ emerging to-day from the barrier
(otherwise stated: crossing to-day the Hubble radius $H_{0}^{-1}$),
namely $k_{0}=a_{0}H_{0}$.

The final number of produced gravitons, for each mode k, is given by
$|c_{-}(k)|^{2}$.
The corresponding energy density $\rho_{g}$, in the proper frequency
interval $d\om$, is obtained by summing over the two polarization
states, and is related to $c_{-}$ by$^{5,7}$
$$
d\rho_{g}=2\om|c_{-}|^{2} 4\pi\om^{2} {d\om\over (2\pi)^{3}} \eqno(3.11)
$$

The spectral energy density $\rho(\om)=\om{d\rho_{g}/ d\om}$, which
is the variable usually adopted$^{3,5-7}$ to characterize the graviton
energy distribution, turns out then to be parametrized as follows
$$
\rho(\om)\simeq\om^{4}(k\eta_{1})^{-2|\mu- \nu|}~~~,~~~~~~~k_{2}<k<k_{1}
$$
$$
\rho(\om)\simeq	\om^{4}(k\eta_{1})^{-2|\mu- \nu|}(k\eta_{2})^{\mp 2
|\sigma- \mu|}~~~~,~~~~k_{0}<k<k_{2} \eqno(3.12)
$$

For later comparison with present  observational data , it is
convenient to replace all comoving frequencies  $k$ by the associated
proper frequency $\om={k/ a(t)}$, and to express the spectral
distribution in terms of the final curvature scale $H_{1}\equiv
H_(\eta_{1})$,
reached at the end of the inflationary phase,
$H_{1}\simeq (a_{1}\eta_{1})
^{-1}=\om_{1}$. Since
in our model $\eta_{1}$ is also the beginning of the
radiation-dominated evolution for $a(t)$, it follows that $H$ can be
expressed in terms of the radiation energy density $\rho_{\ga}$, as
$$
H_{1}^{2}\simeq\left(k_{1}\over a_{1}\right)^{2}\simeq G\rho_{\ga}(
\eta_{1})~~~~,~~~~
G\rho_{\ga}(t)\simeq\left(k_{1}\over a_{1}\right)^{2}\left(a_{1}\over a(
t)\right)^{4} \eqno(3.13)
$$
Note that we have used the Newton constant $G\simeq M_{p}^{-2}$ as the
effective gravitational coupling during the post-inflationary
cosmological evolution; the allowed deviations from this value turn  out
to be indeed negligible for our determination of the spectral behaviour
(see Sec.5).

By using eq.(3.13), and by measuring $\rho(\om)$ in units of critical
energy density $\rho_{c}$, the spectral distribution (3.12) can be
recast finally in the convenient form ($\Om(\om)\equiv{\rho(\om)/
\rho_{c}}$)
$$
\Om(\om,t)\simeq GH_{1}^{2}\Om_{\ga}(t)\left(\om\over\om_{1}\right)^{4-2
|\mu - \nu|}~~~~~~~,~~~~~~\om_{2}<\om<\om_{1}
$$
$$
\Om(\om,t)\simeq GH_{1}^{2}\Om_{\ga}(t)\left(\om\over\om_{1}\right)^{4-2
|\mu - \nu|} \left(\om\over\om_{2}\right)^{\mp 2|\sigma -\mu|}~~~,~~~\om
_{0}<\om<\om_{2} \eqno(3.14)
$$
where $\Om_{\ga}(t)={\rho_{\ga}(t)/\rho_{c}}$ is the fraction of
critical energy density present in  the form of radiation, at the given
observation time $t$.

This spectrum is parametrized by the scale $H_{1}$, and by the
kinematical indices $\mu$, $\nu$, $\sigma$, which determine
its frequency
behaviour. It may be interesting to note that the high frequency part of
the spectrum is decreasing, flat or increasing depending on whether
$|\mu - \nu|$ is larger, equal or smaller than $2$.
For a primordial phase corresponding to isotropic inflation of $d=3$
spatial dimensions, with frozen dilaton and internal radius ($\b=\b_{1}=
\ga=\ga_{1}=0$), eq.(3.4) gives, in particular, $|\mu- \nu|=1+\a$, so
that the behaviour of the spectrum is the same as that of the curvature
scale. For a de Sitter phase ($\a=1$) one recovers indeed the well
known flat spectrum$^{2,23}$ ($\Om\simeq const.$), while
for superinflation ($0<\a<1$) one obtain the growing spectrum recently
discussed in Ref.22.

In the general case in which $d\not=3$, and the additional contributions
 of a dilaton variation (as well as those of dimensional reduction) are
included, however, the spectral behaviour may be flat or decreasing
even if the curvature is growing. What is important to stress
is that, in any case, all observational data and  constraints on the
present background of cosmic gravitational waves can be translated,
thanks to eq.(3.13), into direct information on the curvature scale $H_{
1}$ (marking the transition from the
primordial inflationary phase to the
standard decelerated scenario), and on the kinematics of the background
evolution. This possibility will be discussed in Sec.5.

We conclude this section with an estimate of the transition frequencies
$\om_{1}$ and $\om_{2}$.
At the present time $t_{0}$, the minimal proper frequency $\om_{0}$ is
determined by the to-day value of the Hubble radius, i.e. $\om_{0}=H_{0
}\sim 10^{-18}Hertz$. The frequency $\om_{2}$, corresponding to the
matter-radiation transition,
can be easily related to $\om_{0}$ by noting
that $a(t)\sim t^{2\over 3}$ during the matter-dominated regime, so that
$$
{\om_{2}\over\om_{0}}={k_{2}\over k_{0}}\simeq {H_{2}a_{2}\over H_{0}a_{0
}}\simeq \left({t_{0}\over t_{2}}\right)^{1\over 3}=\left({a_{0}\over a_{2
}}\right)^{1\over 2} \eqno(3.15)
$$
On the other hand, the radiation temperature evolves adiabatically ($aT=
const.$), so that the ratio (3.15) can be expressed in terms of the
temperature $T_{2}$ at the transition time,
$$
{\om_{2}\over\om_{0}}\simeq \left({T_{2}\over T_{0}}\right)^{1\over
2}\sim 10^{2} \eqno(3.16)
$$
where $T_{0}\sim 1^{0} K$ is the present temperature of the radiation
background.

In a similar way we can relate  $\om_{0}$ to the maximal cutoff
frequency $\om_{1}$, which depends on the final curvature scale $H_{1}$.
We can put, in fact,
$$
{\om_{1}\over\om_{0}}={k_{1}\over k_{0}}\simeq{H_{1}a_{1}\over H_{0}a_{0
}}=\left({H_{1}a_{1}\over H_{2}a_{2}}\right)\left({H_{2}a_{2}\over H_{0
}a_{0}}\right) \eqno(3.17)
$$
and we note that, during the radiation dominated evolution, $a\sim t^{1
\over 2}\sim H^{-{1\over 2}}$. We have, moreover, $H_{2}\sim 10^{6}H_{0
}$ and (in units of Planck mass) $H_{0}\sim 10^{-61}M_{p}$; therefore
$$
{\om_{1}\over\om_{0}}\simeq 10^{2}\left({H_{1}\over M_{p}}\right)^{1
\over 2}\left({M_{p}\over H_{0}}\right)^{1\over 2}\left({H_{0}\over H_{2
}}\right)^{1\over 2}\sim 10^{29}\left({H_{1}\over M_{p}}\right)^{1\over
2} \eqno(3.18)
$$

\vskip 2cm

{\bf 4.The squeezing parameter}

Another phenomenological signature of the primordial cosmological
transitions, encoded into the cosmic gravity-wave background, is the
squeezing parameter which characterizes the quantum state of the
gravitons produced from the  vacuum$^{13}$. This parameter
is directly related to the Bogoliubov coefficients, and
is thus sensible to all the various components of the production
process, including a possible variation
of the dilaton background, just like the
spectral energy distribution.

The graviton production discussed in the
previous section is based on the expansion of the gravitational
perturbation in terms of $|in>$ and $|out>$ states, namely
$$
\psi(k,\eta)=b\psi_{in}+ b^{\dagger}\psi_{in}^{\ast}
$$
$$
\psi(k,\eta)=a\psi_{out}+ a^{\dagger}\psi_{out}^{\ast} \eqno(4.1)
$$
for each mode $k$.
The two sets of solutions are connected by a Bogoliubov transformation
which, when expressed in terms of the "in" and "out" mode solutions,
takes the form
$$
\left(\matrix{\psi_{in}\cr
\psi_{in}^{\ast}\cr}\right)=\left(\matrix{c_{+}&c_{-}\cr
c_{-}^{\ast}&c_{+}^{\ast}\cr}\right)\left(\matrix{\psi_{out}\cr
\psi_{out}^{\ast}\cr}\right) \eqno(4.2)
$$
where $c_{\pm}$ are defined, according to eq,(3.4), as $c_{\pm}=
{B_{\pm}/ C}$.
The equivalent relation among the corresponding annihilation and
creation operators of the second-quantization formalism
is then
$$
\left(\matrix{a\cr
a^{\dagger}\cr}\right)=\left(\matrix{c_{+}&c_{-}^{\ast}\cr
c_{-}&c_{+}^{\ast}\cr}\right)\left(\matrix{b\cr
b^{\dagger}\cr}\right)  \eqno(4.2)
$$

If the Bogoliubov transformation is parametrized by two real numbers,
$r$
and $\theta$, in such a way that
$$
c_{+}=\cosh{r}~~~~,~~~c_{-}^{\ast}=-e^{2i\theta}\sinh{r} \eqno(4.4)
$$
the transformation  (4.3) can be rewritten as
$$
a=S^{\dagger}bS~~~,~~~a^{\dagger}=S^{\dagger}b^{\dagger}S  \eqno(4.5)
$$
where $S$ is a unitary operator  defined by
$$
S=\exp{[{1\over 2}z(b^{\dagger})^{2}-{1\over 2}z^{\ast}b^{2}]}
{}~~~~~~,~~~~~z=re^{2i\theta} \eqno(4.6)
$$
This is a so-called "squeezing" operator:
when applied to the vacuum (or, more generally, to a coherent state),
generates a state for which the quantum fluctuations of the operator
$X\sim b+ b^{\dagger}$ (or its canonical conjugate) can be arbitrarily
squeezed for a suitable choice of $r$ (see for instance Ref.(24)).
In particular, $\Delta X\to 0$ for $r\to\infty$.

The cosmic gravitons arising from the background transitions are thus
produced in a squeezed state, with a parameter $r$ which, according to eq
(4.4), is given by
$$
r=\ln{(|c_{-}|+\sqrt{|c_{-}|^{2}+1})}  \eqno(4.7)
$$
According to the model of background evolution considered in the
previous section, and for $\om>\om_{2}$, the relic graviton background
may be characterised, in general, by the following squeezing parameter
$$
\eqalign{r(\om)&\simeq\ln{|c_{-}|}\simeq~
-|\mu -\nu|\ln{\left({\om\over\om_{1
}}\right)} \cr
&\simeq |\mu -\nu|\left[25-\ln{\left({\om\over Hertz}\right)}+{1\over 2
}\ln{\left({H_{1}\over M_{p}}\right)}\right] \cr}\eqno(4.8)
$$
(we have used eq.(3.7) for $c_{-}$, and the estimate (3.18)
for $\om_{1}$).

The first term in eq.(4.8) is expected to be the dominant one, at
least in the range of frequencies accessible,
 in a (hopefully) not too distant future, to a direct observation$^{3
,4}$.
The second term takes into account the variation of $r$ with frequency,
and the third term provides a correction if the transition curvature
scale is different from the Planck scale.
A direct measurement of this parameter, at some definite value of
frequency, would provide then significative information both on the
curvature scale $H_{1}$, and on the background (dilaton included)
dynamical evolution, through the $|\mu- \nu|$ dependence.
\vskip 2cm
{\bf  5.Phenomenological constraints on the graviton spectrum}

The present energy distribution of a cosmic gravity-wave background is
mainly constrained by three kind of direct observations$^{3,4}$:
the absence of fluctuations in the millisecond pulsar-timing data, the
critical density value, and the isotropy of the cosmic microwave
background radiation (CMBR).
The first one applies on a narrow frequency interval around $\om_{p}\sim
10^{-8} Hertz$, while the other  two at all frequencies (the third one
provides a bound which is frequency-dependent).
Their relative importance, and the frequency at which they provide the
must significant constraint, depend on the slope of the graviton energy
spectrum $\Om(\om)$.

For a stochastic graviton background, the bound on the spectrum
following from the CMBR isotropy constrains the wave amplitude $h(\om)$,
and scales like $\om^{-2}$.
It provides then the most significative bound at the minimum frequency
$\om_{0}$ (where it implies$^{3}$ $\Om\le 10^{-8}$), unless we have a
spectrum which in its low frequency band ($\om<\om_{2}$) grows faster
than $\om^{2}$.
If the spectrum is growing at all frequencies, however, the most
significant constraint, for the present values of experimental data, is
provided in any case by the critical density bound $\Om\me 1$, applied
to the highest frequency $\om_{1}$.

According to our three component model of background evolution, the
spectrum may be increasing at low frequencies, and simultaneously flat
or decreasing in the high frequency sector, only if (see eq.(3.14))
$$
\mu > \nu~~~,~~~2\le|\mu -\nu|<2+|\sigma - \mu|  \eqno(5.1)
$$
Even in such a particular case, however, the growth of the low frequency
sector cannot be significantly faster than $\om^{2}$, since, as we shall
see later, $|\sigma- \mu|$ is not allowed to be notably larger than 1 by
the present limits on the variation of the fundamentals constants.
Therefore, the energy distribution of the graviton background can be
significantly constrained by imposing on eq.(3.14) the three following
bounds:
$$
\Om(\om_{1})<\Om_{c}~~~,~~~\Om(\om_{p})<\Om_{p}~~~,~~~\Om(\om_{0})
<\Om_{i}  \eqno(5.2)
$$
where $\om_{p}\sim 10^{-8}Hertz$, and $\Om_{c}$, $\Om_{p}$, $\Om_{i}$
are the present value of the bounds on the energy density imposed,
respectively, by critical density, pulsar timing data, and CMBR isotropy.

For our discussion of the constraints, it may be convenient to
simplify the notations by defining the variables
$$
x=|\mu -\nu|~~~,~~~y=|\sigma -\mu|~~~,~~~z=\log{\left({H_{1}\over
M_{p}}\right)} \eqno(5.3)
$$
By using $\Om_{\ga}(t_{0})\sim 10^{-4}$ for the present critical fraction
of radiation energy density, and by inserting in eq.(5.2) the values of
$\om_{0}$, $\om_{1}$, $\om_{2}$ determined in Sec.3, the three
constraint equations in the parameter space, for our model of background
evolution, can thus be written
$$
z< 2 + {1\over 2}\log{\Om_{c}}
$$
$$
z< {1\over x}(80 + \log{\Om_{p}}) -38
$$
$$
z< {1\over x}(120 \mp 4y + \log{\Om_{i}}) -58 \eqno(5.4)
$$
They follow, respectively, from the critical density, pulsars and
isotropy bounds, and they define an allowed region in the $(x,y,z)$
space which provides information on the past evolution of our Universe.

In order to discuss the extension of this region it should be noted,
first of all, that the range of variation of the variable $y$,
$$
y=|\sigma -\mu|={1\over 2}|n_{1}(\b_{2}-\b_{1})+2\ga_{2}-\ga_{1}-2|
\eqno(5.5)
$$
which parametrizes  the time evolution  of the dilaton and of the
compactification radius during the matter and radiation dominated era
(recall eq.(3.2)), is severely constrained by the present bounds on the
variation of the fundamental constants.

Indeed, in a Brans-Dicke frame, and in a higher dimensional context with
$n=D-4$ dimensions lying in a compact internal space, with scale factor
 $b(t)$, the effective four dimensional Newton constant $G_{N}$ evolves
 in time like $G_{N}\sim {e^{\phi}/ b^{n}}$.
We have then
$$
{\dot G_{N}\over G_{N}}=\dot\phi-n{\dot b\over b}     \eqno(5.6)
$$
During the matter dominated era the variation of the extra spatial
dimensions is constrained by$^{25}$
$$
|{\dot b/ b}|\le 10^{-9}H_{0}\eqno(5.7)
$$
and the variation of $G_{N}$ by$^{26}$
$$
|{\dot G_{N}/ G_{N}}|<10^{-1}H_{0} \eqno(5.8)
$$
where we have taken for $H_{0}$ the largest value allowed
to-day, $H_{0}\simeq 10^{-10} yr^{-1}$.
These two bounds imply $|\dot\phi|<10^{-1}H_{0}$. But, according to our
parametrization (3.2), $\dot\phi=\ga_{2} H$ and ${\dot b/ b}=-\b_{2
}{H/ 2}$. It follows that
$$
|\b_{2}|\le 10^{-9}~~~,~~~|\ga_{2}|<10^{-1}    \eqno(5.9)
$$

Consider now the radiation-dominated era.
During this phase, the best limits on $\dot\phi$ and $\dot b$ are
obtained from the primordial nucleosynthesis.
Denoting by $b_{nucl}$, $G_{nucl}$, and by $b_{0}$, $G_{0}$, the values
of the radius  of the internal space and of the Newton constant, at the
epoch of nucleosynthesis and at the present epoch, respectively, one
obtains that the change of b must be bounded by$^{25,27}$
$$
{b_{nucl}\over b_{0}}=1+\epsilon~~~,~~~|\epsilon|<10^{-2} \eqno(5.10)
$$
while the change of G is constrained by$^{28}$
$$
{G_{nucl}\over G_{0}}=1+\epsilon~~~,~~~|\epsilon|<3\times10^{-1}
\eqno(5.11)
$$

Translated into limits on the time variation of $b$
and $\phi$, according
to the parametrization (3.2), they imply
$$
|\b_{1}|<10^{-3}~~~,~~~|\ga_{1}|<10^{-1}   \eqno(5.12)
$$

The dilaton contribution is thus the dominant source of uncertainty in
the value of the parameter $y$. Even taking into account the maximum
allowed uncertainty, however, it follows from eqs.(5.9) and (5.12) that
$$
0.9\le y \le 1.1 \eqno(5.13)
$$
A first rough evaluation  of the allowed region in the parameter space
is thus obtained by fixing $y=1$ in eqs.(5.4) (the allowed deviation of
$y$ from 1  is too small to be significant in view of our previous
approximations).

We have to insert,
moreover, in eq.(5.4) the values of the bounds implied
by the present experimental data.
We shall put $\Om_{c}=1$ (in order to avoid that the produced gravitons
over-close our present universe), $\Om_{p}=10^{-6}$ as implied (at the
$99 \%$  confidence level) by recent results from pulsar timing$^{29}$,
and $\Om_{i}=10^{-8}$, following from the constraint
$^{30}$ $h<10^{-5}$
 on the gravity-wave amplitude. With this data, the constraint eqs.(5.4)
 become
$$
z<2
$$
$$
z<{74\over x}-38
$$
$$
z<{1\over x}(112 \mp 4) -58   \eqno(5.14)
$$

We recall that the negative (positive) sign in the last equation
corresponds to $\mu<\nu$ ($\mu>\nu$).
It should be mentioned, moreover, that in the context of a more
stringent analysis, the first bound $z<2$ could be replaced by $z<0$,
following from the fact
that early nucleosynthesis seems to imply$^{31}$ , at
high frequency, $\Om<10^{-4}$ for the energy density
distribution of massless particles. This would correspond to a maximum
scale $H_{1}<M_{p}$ instead of $10^{2} M_{p}$. This conclusion is,
however, model dependent, and in this paper we prefer to rely on
constraints following directly from observations.

The allowed region of the $(x,z)$ plane delimited by eqs.(5.14) is
illustrated in  {\bf Fig.1}.
Because of the uncertainty of the experimental data, which has not been
completely taken into account in our discussion, and because of the
approximations made, this figure is expected to give only a qualitative
picture of the phenomenological scenario.
Nevertheless, we can draw from our analysis the following general
conclusions.

1) There is a maximum allowed value for the curvature
scale $H_1$ at the epoch
of the transition from the phase of accelerated expansion, dilaton
growth and dimensional reduction, to the decelerated
radiation-driven evolution,i.e. $H_{1}\le 10^{2}M_{p}$.

2) Models characterised by a sufficiently high scale, $H_{1}\ge10^{-2
}M_{p}$
, are constrained by pulsar timing if $\mu\ge\nu$, and by CMBR isotropy
if $\mu\le\nu$.

3) For any given scale $H_{1}$ lower than the maximum one there is a
limiting slope of the spectrum, below which that scale is forbidden.
Within our approximations, the limiting slope for a scale $H_{1}$
is fixed by
$$
x< {108\over {58+\log{\left({H_{1}\over M_{p}}\right)}}} \eqno(5.15)
$$
if $\mu<\nu$, and
$$
x< {74\over {38+\log{\left({H_{1}\over M_{p}}\right)}}} \eqno(5.16)
$$
if $\mu>\nu$.
In the first case (which corresponds to all the physical models
considered in the next section), the maximum scale $10^{2}M_{p}$ is
allowed for $x\me 1.8$, while the Planck scale can be reached for
$x\me {54/ 29}\simeq 1.86$. A four dimensional inflationary
background, with frozen dilaton and radius of the internal dimensions
($\ga=\b=0$, $d=3$), corresponds in particular to $\mu -\nu=-\a -1 <0$,
and the Planck scale is thus reached for $\a\me{25/ 29}$, in agreement
with the results of a previous analysis$^{22}$.

4) Finally,
models corresponding to a spectrum which is flat or decreasing
at high frequencies
(i.e. with $x\ge 2$), are characterized by a maximum allowed
scale $H_{1}\le 10^{-4}M_{p}$. We thus recover the well known bound on
the scale of a four dimensional de Sitter inflation$^{2,22}$, since in
that case $x=|\a+1|=2$ and one obtains the usual flat spectrum.
\vfill\eject
\centerline{\bf 6. String cosmology pre-big-bang }
\centerline{\bf and other higher-dimensional models}
\vskip 1true cm
In the standard cosmological model the curvature is monotonically
increasing as we go back in time, and blows up at the initial
singularity. A possible classical alternative to the singularity would
seem to be provided by an initial inflationary de Sitter phase, at
constant curvature, which extends in time indefinitely toward the past.
However, as discussed in a recent paper  $^{32}$, eternal exponential
expansion, with no beginning, is impossible in the context of the
conventional inflationary scenario,
so that a primordial phase
of constant curvature does not help to solve the problem of the initial
singularity. Moreover, according
to the  constraints reported in the previous section,
the constant value of the curvature during the initial de Sitter phase
should lie at least four orders of magnitude below the Planck scale;
this may seem unnatural, if one believes that the growth of the curvature
is stopped and that the primordial curvature becomes stable  just
because of quantum effects.

A different alternative has been recently suggested, on the grounds of
string theory motivations$^{10,12,33}$, in which the singularity is
avoided  because the curvature grows up to a maximum (Planckian) value
and then decreases back to zero.
The standard radiation-dominated phase is then preceeded in time by a
phase with "dual" dynamical behaviour (the curvature and the dilaton are
growing, $\dot H >0$, $\dot\phi>0$, the evolution is accelerated, $\ddot
a>0$), called$^{12}$ "pre-big-bang".
Particular examples of such a scenario are thus provided also by
earlier models
of superinflation and dynamical dimensional reduction, discussed in the
context of Kaluza-Klein cosmology$^{34-36}$.

In this section we want to stress that if the initial configuration of
our model of background evolution (i.e. for $\eta< -\eta_{1}$, see
eq(3.11)) corresponds to a pre-big-bang scenario of this type, the
consequent graviton spectrum is always growing fast enough to avoid the
de Sitter bound $H<10^{-4}M_{p}$ (i.e. $x<2$),
and to allow the Universe to inflate
up to the maximal curvature scale, consistently with the bounds of the
previous section.

Consider indeed the perfect-fluid dominated model of Refs.(34) and (35),
which describes superinflation and dimensional decoupling, and belongs
to the class parametrized by eq.(3.1) with $\ga=0$.
One finds, for this model, $\mu- \nu= -{1\over 2}$, and $x=|\mu - \nu|
=0.5$. The model of Ref.(36)
(based on the toroidal compactification of D=11
supergravity), corresponds to $\ga=0$, $\a=0.26$ and $\b=0.22$,
and gives
$\mu -\nu= -0.49<0$.
The model of string-driven inflation of Ref.(33)
has $\ga=0$, $n>10$, and
for $d=3$ it gives $\mu -\nu= {(4-n)/ 3n}<0$.
Finally, a typical pre-big-bang model$^{12}$,
dual to the standard radiation phase, satisfies the Brans-Dicke equations
(2.2) and (2.3) (with $\om=1$) for
$$
\ga=2d~~~,~~~\a={2\over {3+d+n}}=\b    \eqno(6.1)
$$
and implies
$$
\mu -\nu={-2\over {3+d+n}}<0    \eqno(6.2)
$$
For all these models we have $\mu -\nu<0$, and $|\mu -\nu|<1.8$ (for any
allowed number of internal dimensions), so that their final curvature
scale is only constrained by the closure density bound.

We want to comment, finally, on the possibility that the CMBR anisotropy
recently measured$^{37}$ by COBE be at least partially
determined, at the
quadrupole level, by a cosmic graviton background.
It has been already pointed out$^{38}$,
indeed, that a stochastic background of
gravitational waves with flat spectrum, generated by a primordial
de Sitter inflationary phase, could produce the entire observed signal,
provided de Sitter inflation occurred
at a vacuum energy scale  $M_Pv^{1/4}\simeq 1.5\times 10^{16}$ GeV
(at the $95\%$ confidence level). This translates into a value of the
Hubble constant $H=(8\pi M_P/3)v^{1/2}\sim 10^{-5}M_P$, which is not in
conflict with the previously reported bound
($H_{1}\me 10^{-4}$ for $x=2$, see {\bf Fig.1}).

It should be noted, however, that a four-dimensional de Sitter inflation
is not the only primordial phase which can be associated to a flat
graviton spectrum. Indeed, in a more general
higher-dimensional Brans-Dicke scenario,
all the models  with $|\mu- \nu|=2$ provide a flat high
frequency spectrum. Included in this class, in particular, are all the
$(d+1)$-dimensional models providing a phase with variable dilaton and
isotropic superinflationary expansion, characterized in conformal time
(according to eq.(3.1)) by the power $\a={2/ {(d-1-\ga)}}$.

It remains still open, moreover, the interesting possibility that the
COBE anisotropy may be fitted a by
non-flat graviton spectrum$^{39}$ with,
in particular, $x<2$, as predicted by the string pre-big-bang models.
In this case we may expect, according to {\bf Fig.1},
that the COBE data  will
select a higher transition scale $H_{1}$, and in such a case they could
be  interpreted, instead of a first direct evidence, via gravitational
waves, for the GUT scenario$^{38}$, as evidence
for the dilaton-driven string cosmology scenario. In order to
discriminate between these two (exciting) possible interpretations,
however, one should try to probe directly the energy density of the
cosmic graviton background at some given frequency,
for example through
a gravity wave detector (such as LIGO$^{4}$), or by means of
astrophysical methods (such
as timing measurements of millisecond pulsars$
^{29}$).

\vskip 2cm

{\bf 7. Conclusions}

In this paper we have considered a three-component model of cosmological
evolution in which the standard radiation and matter-dominated
expansion
of the three-dimensional space is preceeded in time by a
general $d$-dimensional
phase of accelerated (i.e. inflationary) expansion.
We have included, moreover, a possible variation of the effective
gravitational coupling and of the compactification scale, parametrized,
respectively by a logarithmic time dependence of the dilaton field, and
by a power law evolution of the internal scale factor.

We have shown that the linearised equation for a metric fluctuation,
obtained by perturbing the Brans-Dicke equations around this background,
contains a coupling  of the perturbation to the background metric and to
the dilaton field $\phi$. As a consequence, both the dimensional
reduction process and the variation of $G$ (via $\dot\phi$) contribute
 (besides inflation) to the process of the amplification of the
 gravitational perturbations (i.e. to the graviton production).

We have computed the spectral distribution $\Om(\om)$ of the energy
density stored to-day in a cosmic graviton background (and the
associated squeezing parameter $r(\om)$), taking  into account all
possible contributions. The frequency behaviour of the spectrum turns
out to be clearly related to the temporal behaviour of the background
fields ($g_{\mu\nu}$ and $\phi$); the observational constraints on $\Om(
\om)$ provide then significative information both on the kinematics of
the background evolution, and on the curvature scale $H_{1}$
characterizing the transition from the primordial inflationary phase
(with variable dilaton), and the standard radiation-dominated phase.

We have shown, in particular, that for flat or decreasing spectra the
transition scale cannot overcome a maximum value which lies, typically,
four orders of magnitude below the Planck scale.
For growing spectra,
on the contrary, the allowed transition scale can be
as high as the Planck one (and somewhat higher).

We have stressed, finally, that the contribution of the dilaton
background to the cosmic production of gravitons may simulate the usual
flat four-dimensional de Sitter spectrum, even if the inflationary
evolution of the scale factor is not of the exponential type, and the
curvature scale is growing, instead of constant, during the inflation.
As a consequence, one could try to interpret the recently measured COBE
anisotropy not only as evidence for de Sitter inflation at the GUT
scale$^{38}$, but also (alternatively) as a possible evidence for a
dilaton-driven string cosmology scenario$^{10,12}$.
\vskip 2cm
{\bf Acknowledgments}

We are very grateful to G.Veneziano for many useful discussions.

\vskip 2 cm
\centerline{\bf References}

\item{1.} L.P.Grishchuk, Sov.Phys.JETP 40,409(1975);

 L.P.Grishchuk and A.Polnarev, in"General Relativity and Gravitation",

 ed. by A.Held (Plenum, New York, 1980) ol.2, p.393

\item{2.}V.A.Rubakov, M.V.Sazhin and A.V.Veryaskin, Phys.Lett.B115,

189(1992);

R.Fabbri and M.D.Pollock, Phys.Lett.B125,445(1983);

L.F.Abbott and M.B.Wise, Nucl.Phys.B244,541(1984)

\item{3.}L.P.Grishchuk, Sov.Phys.Usp.31,940(1988);

Sov.Sci.Rev.E.Astrophys.Space Phys.7,267(1988).

\item{4.}K.S.Thorne, in"300Years of Gravitation", ed.by S.W.Hawking and

W.Israel (Cambridge Univ.Press,Cambridge 1988) p.330

\item{5.}B.Allen, Phys.Rev.D37,2078(1988)

\item{6.}V.Shani, Phys.Rev.D42,453(1990)

\item{7.}L.P.Grishchuk and M.Solokhin, Phys.Rev.D43,2566(1991)

\item{8.}J.Garriga and E.Verdaguer, Phys.Rev.D39,1072(1991)

\item{9.}M.Demianski, in Proc. of the 9th Italian Conference on General

Relativity and Gravitational Physics
(Capri,1990), ed. by R.Cianci et al.,

(World Scientific, Singapore, 1991) p.19;

L.Amendola, M.Litterio and F.Occhionero, Phys.Lett.B237,348(1990);

M.Gasperini and M.Giovannini, Class.Quantum Grav.9,L137(1992)

\item{10.}G.Veneziano, Phys.Lett.B265,287(1991);

M.Gasperini, J.Maharana and G.Veneziano, Phys.Lett.B272,277(1991);

M.Gasperini and G.Veneziano, Phys.Lett.B277,256(1992)

\item{11.}A.A.Tseytlin and C.Vafa, Nucl.Phys.B372,443(1992);

A.A.Tseytlin, Class.Quantum Grav.9,979(1992)

\item{12.}M.Gasperini and G.Veneziano,"Pre-big-bang in string cosmology",

CERN-TH. 6572/92 (July 1992)

\item{13.}L.P.Grishchuk and Y.V.Sidorov, Class.Quantum Grav.6,L161(1989);

Phys.Rev.D42,3413(1990);

L.P.Grishchuk, in Proc.of the Workshop"Squeezed states and

uncertainty relations", University of Maryland, ed. by D.Han, Y.S.Kim

and W.W.Zachary (Nasa Conference Pub.N.31353, 1992) p.329

\item{14.}E.M.Lifshitz, Zh.Eksp.Teor.Phys.16,587(1946);

E.M.Lifshitz and I.M.Khalatnikov, Adv.of Phys.12,208(63)

\item{15.}L.H.Ford and L.Parker, Phys.Rev.D16,1601(1977)

\item{16.}K.A.Meissner and G.Veneziano, Phys.Lett.B267,33(1991);

Mod.Phys.Lett.A6,3397(1991)

\item{17.}L.D.Landau and E.M.Lifshitz, Teoria dei Campi (Editori Riuniti,

Roma, 1985) \S 115, p.486

\item{18.}B.L.Hu, Phys.Rev.D18,969(1978)

\item{19.}M.Abramowitz and I.A.Stegun,
"Handbook of mathematical functions"

(Dover, New York, 1972)

\item{20.}N.D.Birrel and P.C.W.Davies, "Quantum fields in curved space"

(Cambridge Univ.Press, Cambridge, 1982)

\item{21.}E.Verdaguer, "Gravitational particle creation in the early

Universe" UAB-FT-276(November 1991)

\item{22.}M.Gasperini and M.Giovannini, Phys.Lett.B282,36(1992)

\item{23.}A.A.Starobynski, JETP Lett.30,682(1979)

\item{24.}J.Grochmalicki and M.Lewnstein, Phys.Rep.208,189(1991)

\item{25.}J.D.Barrow, Phys.Rev.D.35,1805(1987)

\item{26.}R.W.Hellings, in Proc.of the 10th Course of the Int.School

of Cosmology and Gravitation (Erice, 1987),ed. by V.De Sabbata

and V.N.Melnikov (Kluwer Acad.Pub.,Dordrecht,1988) p.215

\item{27.}E.W.Kolb, M.J.Perry and T.P.Walker, Phys.Rev.D33,869(1986)

\item{28.}F.S.Accetta, L.M.Krauss and P.Romanelli, Phys.Lett.B248,

146(1990)

\item{29.}D.R.Stinebring et al., Phys.Rev.Lett.65,285(1990)

\item{30.}G.F.Smoot, in Proc.of the First Course of the Int. School of

Astrophysics D.Chalonge, (Erice,1991) ed. by N.Sanchez (World Scientific,

Singapore)

\item{31.}V.F.Schwartzaman, JETP Lett.9,184(1969)

\item{32.}A.Vilenkin,"Did the Universe have a beginning?",

CALT-68-1772(1992)

\item{33.}M.Gasperini, N.Sanchez and G.Veneziano,
Nucl.Phys.B364,365(1991)

\item{34.}R.B.Abbott, S.M.Barr and S.D.Ellis, Phys.Rev.D30,720(1984)

\item{35.}D.Shadev, Phys.Rev.D39,3155(1989);

Phys.Rev.D30,2495(1984)

\item{36.}R.G.Morhouse and J.Nixon, Nucl.Phys.B261,172(1985)

\item{37.}G.F.Smoot et al.,"Structure with COBE DMR first Year Map",

 COBE Preprint (21 April 1992)

\item{38.}L.M.Krauss and M.White, "Grand Unification,
gravitational waves, and

 the cosmic microwave background anisotropy", YCTP-P15-92 (April 1992)

\item{39.}T.Souradeep and V.Sanhi, "Density perturbations, gravity waves
and the cosmic microwave background", IUCAA Preprint (July 1992);

F.Lucchin, S.Matarrese and S.Mollerach, "The gravitational wave

contribution to CMB anisotropies and the amplitude of mass fluctuations

from COBE results", Fermilab-Pub-92/185-A (July 1982);

M.Gasperini and M.Giovannini, in preparation

\vfill\eject
\topinsert
\vskip 2 cm
\endinsert
\centerline{\bf Figure caption}
\vskip 1cm
\noi
{\bf Fig.1}

\noi
The maximum allowed value of the transition scale $H_{1}$ (in units of
Planck mass), versus the parameters determining the kinematics of the
background evolution. The allowed region with $H_{1}\me 10^{2}M_{p}$
extends from $x=1.8$ down to $x=0$.
\end